\documentclass[letterpaper, 10 pt, conference]{ieeeconf}  

\usepackage{graphicx} 					
\usepackage{amsmath}
\usepackage{fancyhdr,graphicx,amsmath,amssymb}
\usepackage[ruled,vlined]{algorithm2e}

\SetKwInput{KwInput}{Input}                
\SetKwInput{KwOutput}{Output}              

\title{\LARGE \bf
Correntropy based Robust Decomposition of Neuromodulations
}

\author{Shailaja Akella and Jose C. Principe}

\begin{document}

\maketitle
\thispagestyle{empty}
\pagestyle{empty}

\begin{abstract}
Neuromodulations as observed in the extracellular electrical potential  recordings obtained from Electroencephalograms (EEG) manifest as organized, transient patterns that differ statistically from their featureless noisy background. Leveraging on this statistical dissimilarity, we propose a  non-iterative robust classification algorithm to isolate, in time, these neuromodulations from the temporally disorganized but structured background activity while simultaneously incorporating temporal sparsity of the events. Specifically, we exploit the ability of correntropy to asses higher - order moments as well as imply the degree of similarity between two random variables in the joint space regulated by the kernel bandwidth. We test our algorithm on DREAMS Sleep Spindle Database and further elaborate on the hyperparameters introduced. Finally, we compare the performance of the algorithm with two algorithms designed on similar ideas; one of which is a quick, simple norm based technique while the other parallels the state-of-the-art Robust Principal Component Analysis (RPCA) to achieve classification. The algorithm is able to match the performance of the state-of-the-art techniques while saving tremendously on computation time and complexity.
\end{abstract}

\section{Introduction}
The non - stationarity property of extracellular electrical potentials is a direct consequence of the highly dynamical nature of transitions in the brain as it alternates between complex unpredictable chaos to predictable oscillatory stages [1]. Diffused activity between local neural populations over an extended range in time and space then emerge as the organized, transient patterns called neuromodulations. These manifestations represent the synchronized effort of neural assemblies to process external and internal stimuli, thus encoding significant information regarding underlying pathophysiological processes. Thus, neuromodulations as biological markers have played an important role in the field of Neuroengineering, time and again being useful in sleep analysis [10], Brain Machine Interfaces [11], epilepsy studies [12] and neurorehabilitation [13], among other applications. 
\par
The highly complex unpredictable chaotic stages then corresponds to the featureless background activity which is known to be characterized by a $1/f$ power spectrum. This nature of the noise component is attributed to self - organized criticality that defines the complex $1/f$ state as a metastable state that falls between predictable oscillations and unpredictable, temporally disorganized chaos [1]. Further bolstering this two - component hypothesis, Freeman [2] experimentally confirmed that the probability density function (pdf) of the electrocortigraphy (ECoG) traces from the cat's olfactory bulb conformed to a Gaussian distribution during rest stages while devaited significantly from Gaussianity during the active stages. The goal of signal processing is then to employ appropriate statistical models to classify these components in time while accounting for the highly dynamical nature of the brain states.    
\par 
A simple attempt towards such classification inludes simple supervised two - class classification [3] which necessitates strict control on parameters. On the other hand, state - space methods [4] that use linear models may not be able to capture the non - stationarities in the traces efficiently leading to inherent mixing of sources which would, further, be difficult to identify. Similar mixing of sources is inevitable in methods employing subspace analysis [5]. Among other methods based on the two - component hypothesis, a quick and simple $l^{2}$- norm based technique, called the embedding transform [6], although effective, does not exploit complete time information and therefore, while the model gains in computation time and complexity, it lacks in accuracy. Lastly, the classification obtained by the more complex methods like RPCA [7] is more accurate, however, the long computation times and high complexity makes them unsuitable for high - dimensional settings or long durations of EEG traces.        
\par
In an effort to avoid the above stated pit - falls, we propose a novel non-iterative classification technique which meticulously exploits all time information while simultaneously accounting for computation time and complexity. Further, we also keep in mind the neurophysiologically sound and experimentally valid two - component hypothesis and derive our technique by generalizing results from studies by Walter J. Freeman that elaborate on the departure of neuronal waves from Gaussianity during active stages. We use correntropy as the discriminating metric, that is able to map the differences in the distributions of the two-components of bandpassed EEG traces. The result is a simple vector representation that classifies all observations into the said two classes based on a threshold. We present our results as tested on the DREAMS Sleep Spindle database and further compare the model's performance with the embedding transform technique and the more complex RPCA. The results obtained place the performance of the algorithm at par with the state-of-the-art techniques while simultaneously saving on computation time and complexity. The purpose of the paper is to assess the accuracy of the model and hence, we do not address any medical conclusions. 

\section{Correntropy Based Decomposition}
In this section, we set up the problem statement and provide a description of our classification model. Let $\widetilde {x}\left[n\right] $ be a bandpassed single - channel, single-trial recording, then complying to the two - component hypothesis,  $\widetilde {x}\left[n\right]$ can be decomposed into two sequences:

\begin{equation} \label{eq:1}
\widetilde {x}\left[n\right] = l\left[n\right] + s\left[n\right] 
\end{equation}

where  $l\left[n\right]$ is the temporally-dense, structured spontaneous EEG component and $s\left[n\right]$ is the sparse component comprising only the neuromodulations. A principled model would then incorporate these constraints along with Freeman's findings regarding the statistical differences between these components. Deriving support from matrices, let $\textbf{X}$ be the matrix version of $\widetilde {x}\left[n\right]$ such that the columns correspond to non-overlapping, W-long consecutive segments from $\widetilde {x}$. Similarly, let $\textbf{L}$ and $\textbf{S}$ be the matrix versions of  $ l\left[n\right] $ and $ s\left[n\right] $, respectively where $\textbf{L}$ is a low - rank matrix pertaining to the stationarity of background activity and $\textbf{S}$ is a sparse matrix pertaining to transiently occurring neuromodulations. Therefore,

\begin{equation}\label{eq:2}
\boldsymbol{X} = \boldsymbol{L} + \boldsymbol{S} 
\end{equation}

\par The attempt now is to decompose the matrix $\textbf{X}$ into two matrices, $\textbf{L}$ and $\textbf{S}$, wherein we assume that no portion of the columns of $\textbf{L}$ are affected by $\textbf{S}$. This makes sense as according to the component based hypothesis,  $l\left[n\right]$ and $s\left[n\right]$ have different statistical properties.    
\par 
Correntropy [8] is a pdf based measure of similarity between two random variables controlled by the kernel bandwidth. It derives its root from Information Theoretic Learning where  the non - linearity introduced in the form of a kernel provides access to every even order moment of the joint PDF of the two random variables. Given N data points $\left\{(x_{i},y_{i})\right\}_{i = 1}^{N}$, an estimate of correntropy is given by, 

\begin{equation}\label{eq:3}
\hat {V}_{N,\sigma}(X,Y) = \frac{1}{N} \sum_{i = 1}^{N} \kappa_{\sigma}(x_{i} - y_{i}).
\end{equation}

Using this definition of correntropy and the matrices $\textbf{X}$, $\textbf{L}$ and $\textbf{S}$ as defined in (2), we design the matrix, $\textbf{C}$ and the vector, $Z$, as follows,

\begin{equation}\label{eq:4}
\boldsymbol{{C}_{j,k}} =  \frac{1}{W} \sum_{i = 1}^{W}\kappa_{\sigma}(x_{i,j} - x_{i,k}).
\end{equation}

\begin{equation}\label{eq:5}
{Z}_{j} =  \sum_{k = 1}^{N} \boldsymbol{C_{j,k}}
\end{equation}

where $x_{i,j}$ corresponds to the value at $i^{th}$ row and $j^{th}$ column of the matrix $\textbf{X}$ and $\kappa_\sigma$ is the Gaussian kernel of kernel width, $\sigma$. Essentially, we calculate the correntropy between every pair of W-long segments from matrix $\textbf{X}$, which generates the symmetric matrix, $C$, of order $\left\lfloor\dfrac{N}{W}\right\rfloor  X \left\lfloor\dfrac{N}{W}\right\rfloor$ where $\textbf{$C_{j,k}$}$ represents the correntropy measure, i.e., the degree of similarity between the distributions of columns $x_{(:,j)}$ and $x_{(:,k)}$. Finally, we calculate the column sum of the matrix $\textbf{C}$ which generates the vector $Z$. Here, $Z_j$ is a measure defining the similarity between the $j_{th}$ column of  $\textbf{X}$ with the rest of the signal. 

\par
The temporally dense, stationary background activity would then corresponds to higher values of $Z$; and moreover, given their non - stationary and sparse properties, neuromodulations would map to the lower values of $Z$.  In this way, we cluster the background activity, $\textbf{L}$, and the neuromodulations, $\textbf{S}$, using a simple vector representation exploiting the distribution of the data itself while simultaneously, incorporating information from its time structure. As a final step, we use the skewness measure of $\textbf{L}$ as a test for Gaussianity and a percentile based analysis to extract the matrices $\textbf{L}$ and $\textbf{S}$, and the threshold, $\gamma$ defined as the minimum norm of the sparse neuromodulations. Other measures of Gaussianity may also be used, however, the effectiveness of each is yet to be analysed. Algorithm 1 details the implementation of the method where N is the total length of the EEG trace, M is the duration of each neuromodulation and the other variables are as defined in the section. 
   
\begin{algorithm}
\SetAlgoLined
\KwInput{$ \textbf{X}, \sigma $, M}
\KwOutput{$ \textbf{L}, \textbf{S}, \gamma$}
 \For{$i\gets1$ \KwTo $\left\lfloor\dfrac{N}{W}\right\rfloor$ }{
  	\For{$j\gets1$ \KwTo $\left\lfloor\dfrac{N}{W}\right\rfloor$ }{
		$ Z_i =  Z_i +  \frac{1}{W} \sum_{i = 1}^{W}\kappa_{\sigma}(x_{:,i} - x_{:,j}) $
		}
	}
\For{$\rho\gets1$ \KwTo r}{
$I_L = arg (Z > percentile (Z, \rho))$ \\
$\boldsymbol{L_\rho} = [x_{(:,j)}] $   $ j \in I_L, x_{(:,j)} \in \boldsymbol{X} $  \\
$s_\rho = skewness(\boldsymbol{L_\rho})$ \\

}
$\rho^{*} = arg min (|s_\rho|) $ \\
$\boldsymbol{L} = \boldsymbol{L_{\rho^{*}}}$\\
$I_S = arg (Z < percentile (Z, \rho^{*}))$ \\
$\boldsymbol{S} =  [x_{(:,j)}] $   $ j \in I_S, x_{(:,j)} \in \boldsymbol{X} $  \\
$P = findSnippets(\boldsymbol{S},M) $ \\
$\gamma = min(norm(p_i)) $  $ p_i \in P$ \\
 \caption{Correntropy based Decomposition} 
\end{algorithm}

\section{Methods}

The model was tested on publicly available DREAMS Sleep Spindle Database of University of MONS – TCTS Laboratory and Universitѐ Libre de Bruxelles – CHU de Charleroi Sleep Laboratory. The data comprises of 30-minute-long EEG recordings from 6 patients extracted from whole night polysomnographic recordings. The recordings were all upsampled to 200 Hz for consistency. These excerpts have been scored by two experts which we use as the ground truth for our analysis.  Throughout the paper, W for each subject was chosen as the segment length corresponding to the maximum threshold value calculated and M was chosen as 150 samples. Interestingly, the segment length corresponding to the optimal threshold value was always $\approx M \pm 25 $, except for Subject 4 for which W was equal to 105. The data was normalized for correntropy calculation and lastly, the kernel width, $\sigma$, was set as $\frac{\sigma^{*}} {1.5}$ where $\sigma^{*}$ corresponds to the value given by Silverman's rule [9] for the normalized data.

\subsection{Hyperparameter Analysis:}
Prima facie, the model has 3 main hyperparameters -  W, the segment length considered for correntropy calculation; M, maximum duration of neuromodulations; and $\sigma$, width of the kernel implemented. However, the length of each neuromodulation depends on the neurophysiological principles of the brain region under study, visual inspection, previous studies and appropriate time domain analysis. Further, it is important to note that this hyperparameter will not effect the decomposition in any way and is only needed to calculate an appropriate threshold, $\gamma$. The matrices $\textbf{S}$ and $\textbf{L}$ can be extracted regardless of the value of M. 

\par The kernel width has an important implication in the calculation of correntropy in that, it controls the size of the joint space in which the similarity is measured and assessed. Correntropy is, thus, a localized similarity measure which evaluates similarity in a chosen range of the joint space. For the current model, a value smaller than the one obtained by Silverman's rule [9] is ideal as it allows for better differentiation between the two classes. However, a value too small would generate meaningless results and must be avoided. 

\par Finally, the optimal values of W and M are closely related, as the correntropy calculated between W-long segments where W $\approx$ M, captures appropriate similarity in the distribution of the two segments, especially, for segments corresponding to neuromodulations. This is obvious from Fig.1 where threshold values were calculated for different segment lengths averaged across all subjects. It can be seen that the threshold values are higher in the range 100 $\leq$ W $\leq$ 180 and the mean duration of spindles as scored by Scorer 1 was found to be 162. However, Scorer 2 happens to select a value of 200 for all spindle durations, we believe that this value corresponds to the maximum spindle duration. This fits with our analysis as from Fig.1, it can be seen that all threshold values after W = 200 are too small to be optimal. 

\begin{figure}
  \includegraphics[width=\linewidth]{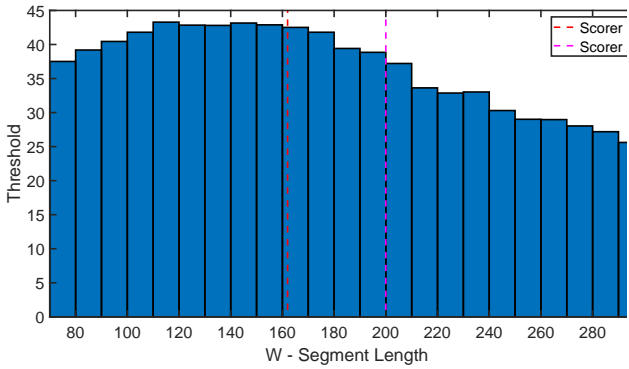}
  \caption{Thresholds calculated for different values of W averaged across all subjects. Dashed lines: Mean duration of spindles as scored by the experts.}
  \label{fig:S1}
\end{figure}   

\subsection{Similarity Vector, $Z$}
The similarity vector, $Z$, represents the classficiation as performed by the model where the neuromodulations are mapped to smaller values while observations corresponding to background activity are mapped to higher values. Some typical vector representations for different segment lengths have been plotted in Fig.2 after sorting the values in the descending order. The typical characteristics of the plot are similar for different segment lengths, however, too long segment lengths can lead to extreme mixing of the two classes while too short segment lengths may not provide correntopy calculation with enough data points to capture similarities in the joint space. Too small segment lengths also limit the kernel width, generating meaningless results. Finally, Fig.3 presents some exemplar snippets as extracted from the cluster corresponding to neuromodulations. 

\begin{figure}
  \includegraphics[width=\linewidth]{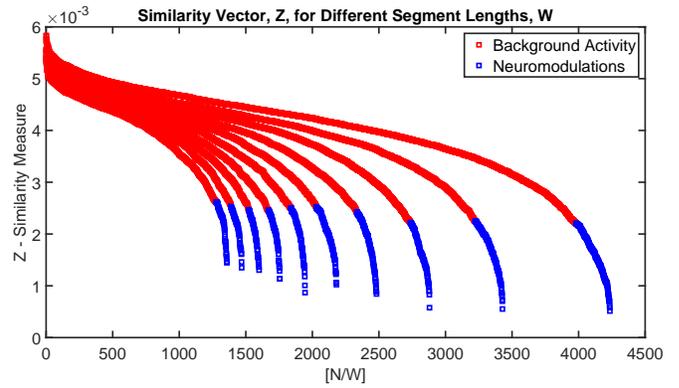}
  \caption{Similarity vectors calculated for different segment lengths,W for Subject 6. Curve on the extreme right corresponds to the smallest segment length and the length increases as we move left. Each curve has been color coded to represent the observations clustered as neuromodulations (blue) and background activity (red).}
  \label{fig:S2}
\end{figure}   

\begin{figure}
  \includegraphics[width=\linewidth]{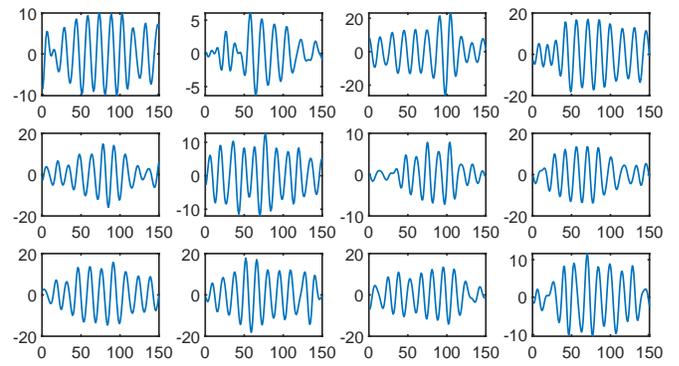}
  \caption{Sample neuromodulation snippets as classified by the model run on bandpass filtered (11 - 16 Hz) recordings of Subject 3.}
  \label{fig:S3}
\end{figure}   

\section{Results}
This section compares the performance of the current model with other state-of-the-art techniques in order to validate its applicability. The two algorithms considered for comparison are: The Embedded Transform [6] and RPCA [7], both of which were derived from similar ideas based on studies by Freeman.  The embedding transform essentially draws an $l^2$ - norm based map in order to emphasize the modulated patterns. In doing so, the background activity accumulates around the main lobe of the distribution while the neuromodulations are pushed to the tail. On the other hand, RPCA is an iterative algorithm that employs convex optimization techniques to minimize the nuclear norm of $\textbf{L}$ while simultaneously minimizing the $l_1$ norm of the sparse matrix, \textbf{S}. This is different from our model in the sense that RPCA is an element - wise model wherein $\textbf{S}$ assumes an arbitrary support and hence, its non - zero elements may affect the columns of $\textbf{L}$. 

\subsection{Threshold Estimation}
We define threshold as the minimum norm of all detected neuromodulations such that any modulation pattern detected with a norm greater than the threshold is a putative phasic event. Physiologically, the threshold tracks the dynamical alternation of the brain states between the complex chaotic and the oscillatory stages. We calculated threshold values for both methods using the technique suggested in the papers [6],[7] and compared those to the values as obtained by our model. Fig. 4 summarizes the threshold values and it can be seen that the thresholds as estimated by our model almost approximate those estimated by RPCA reflecting the accuracy of our model, especially for subjects 5 and 6, where the values essentially overlap.  
   
\begin{figure}
  \includegraphics[width=\linewidth]{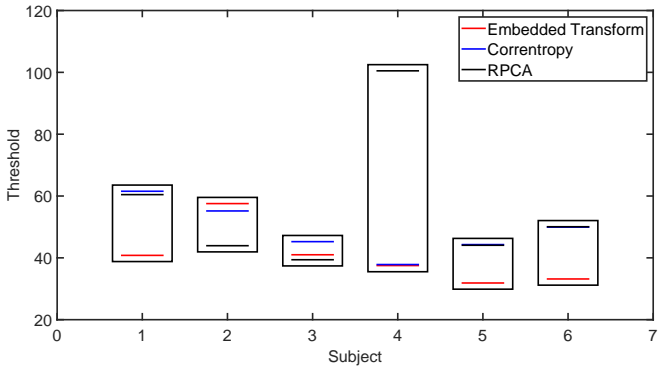}
  \caption{Threshold values as calculated by Embedding Transfom (red), Correntropy Model (blue) and RPCA (black).}
  \label{fig:S4}
\end{figure} 

\subsection{Computation Time}
Two of the major advantages of our model are: firstly, its non - iterative property which helps save tremendously on computation time; second, its explicit use of time structure of the data which simulatneously allows for efficient classification. The computation times of the RPCA method were noted for each subject and compared against the time taken by the correntropy method. The values are as summarized in Table 1. The time taken by the model is significantly less than that taken by RPCA while at the same time, the thresholds estimated by both models are very similar.

\begin{table}[h]
\caption{Computation Time for the Model as Compared to RPCA}
\label{Table 1}
\begin{center}
\renewcommand{\arraystretch}{1}
\begin{tabular}{|c|c|c|}
\hline
Subject & \textit{Correntropy Model (mins)} & \textit{RPCA (mins)} \\
\hline
1& 0.23	& 30.7\\
\hline
2&	0.15	& 35.2\\
\hline
3&	0.22	& 75.4\\
\hline
4&	0.26	& 6.18\\
\hline
5&	0.18 &	73.83\\
\hline
6&	0.18 &	23.95\\
\hline

\end{tabular}
\end{center}
\end{table}

\par All in all, the model's performance can be seen to be at par with the other state-of-the-art methods and it comes with an added advantage of less computational complexity. With 3 hyperparameters which do not require much analysis, the results of the model are promising and have great potential in several biomedical applications. Moreover, being a simple window based method qualifies the model for online implementation.  

\section{Conclusion}
Through this paper, we put forward a simple yet effective, non - iterative, robust classification model to codify non - stationarities of bandpassed single trial, single-channel EEG traces. The method employs information theoretic principles to encode the degree of similarity between time series data. The hyperparameters of the model are easily interpretable. Futher, we successfully demonstrate the ability of the algorithm to achieve results that match the more complex and advanced algorithms in much shorter run times placing the model at par with state-of-the-art methods.

\addtolength{\textheight}{-12cm}

\end{document}